\documentclass[preprint,letterpaper,showpacs]{revtex4}
\usepackage{graphicx}
\def \ds {\displaystyle}
\def \pt {\delta }
\def \al {\alpha }
\def \bal {\bar{\alpha }}
\def \hm {{\bf m}}
\def \Tr {{\rm Tr}}
\def \TC {T_{\rm C}}
\def \bH {{\bf H}}
\def \bz {{\bf z}}

\def \bs {{\bf \sigma }}
\def \vS {{\bf S}}

\def \seff {A_{\rm eff}}
\def \speff {S_{\rm eff}}
\def \uI {\underline{I}}
\def \uG {\underline{G}}
\def \uK {\underline{K}}
\def \vs {{\bf s}}
\def \sgn {{\rm sgn}}
\def \bc {\bar c}
\def \JH {J_{\rm H}}
\def \tJH {\tilde J_{\rm H}}
\def \G0 {\Gamma^{(0)}}
\begin{document}

\title{Thermodynamic Consistency of the Dynamical Mean-Field Theory of the Double-Exchange Model}
\author{Randy S. Fishman$^*$, Juana Moreno$^{\dagger }$, Thomas Maier$^{\bullet}$, and Mark Jarrell$^{\ddagger }$}
\affiliation{$^*$Condensed Matter Sciences Division, Oak Ridge National Laboratory, Oak Ridge, TN 37831-6032}
\affiliation{$^{\dagger }$Physics Department, University of North Dakota, Grand Forks, ND 58202-7129}
\affiliation{$^{\bullet}$Computer Science and Mathematics Division, Oak Ridge National Laboratory, Oak Ridge, TN 37831-6032}
\affiliation{$^{\ddagger }$Deparment of Physics, University of Cincinnati, Cincinnati, OH 45221}

\begin{abstract}

Although diagrammatic perturbation theory fails for the dynamical-mean field theory 
of the double-exchange model, the theory is nevertheless $\Phi $-derivable and hence thermodynamically
consistent, meaning that the same thermodynamic properties are obtained from either the partition function 
or the Green's function.  We verify this consistency by evaluating the magnetic susceptibility and 
Curie temperature for any Hund's coupling.

\end{abstract}
\pacs{75.40.Cx, 75.47.Gk, 75.30.-m}

\maketitle

The dynamical mean-field theory (DMFT) formulated in the late 1980's by M\"uller-Hartmann \cite{mul:89}
and Metzner and Vollhardt \cite{met:89} has developed into one of the most powerful many-body techniques for studying 
electronic models such as the Hubbard \cite{fre:95,geo:96} and double-exchange (DE) \cite{fur:95,mil:96,aus:01,fis:03,che:03} 
models.  This theory is believed to become exact in the limit of infinite dimensions and to capture the physics 
of correlated electron systems even in three dimensions.  Recent work on dilute magnetic semiconductors has
used DMFT to study variants of the DE model \cite{cha:01,ara:04} with less than one local moment per site.  
In this paper, we reach the surprising conclusion that, unlike for the DMFT of the Hubbard model \cite{geo:96},
a diagrammatic perturbation theory containing only electronic degrees of freedom
fails for the DMFT of the DE model.  Nevertheless, we show that the theory remains $\Phi $-derivable 
in a more restrictive sense, which still implies that the 
partition function and Green's function produce consistent results for thermodynamic properties 
such as the magnetic susceptibility and Curie temperature. 
   
The Hamiltonian of the DE model is given by
\begin{equation}
\label{ham}
H=-t\sum_{\langle i,j \rangle }\Bigl( c^{\dagger }_{i\al }c^{\, }_{j\al }
+c^{\dagger }_{j\al }c^{\, }_{i\al } \Bigr) -2 \JH \sum_i \vs_i \cdot \vS_i
\end{equation}
where $c^{\dagger }_{i\al }$ and $c_{i\al }$ are the creation and destruction operators 
for an electron with spin $\alpha $ at site $i$, 
$\vs_i =(1/2) c^{\dagger }_{i\al } \bs_{\al \beta } c^{\, }_{i\beta }$
is the electronic spin, and $\vS_i=S\hm_i $ is the spin of the local moment.  
Repeated spin indices are summed.  Within DMFT, the effective action on site 0 
above $\TC$ in zero field is given by
\begin{equation}
\label{act}
\seff (\hm) =-T\sum_n\bc_{0\al }(i\nu_n) \Bigl\{ G_0(i\nu_n)^{-1}\delta_{\al \beta }
+ \tJH \bs_{\al \beta }\cdot \hm\Bigr\} c_{0\beta }(i\nu_n), 
\end{equation}
where $\tJH =\JH S$, $\nu_n=(2n+1)\pi T$, $\bc_{0\al }(i\nu_n)$ and $c_{0\al }(i\nu_n)$ are 
now anticommuting Grassman variables, and $G_0(i\nu_n)$ is the bare Green's function containing dynamical 
information about the hopping of electrons from other sites onto site 0.

Because $\seff (\hm )$ is quadratic in the Grassman variables, the full Green's function 
$G(i\nu_n)_{\al \beta }$ at site 0 may be readily solved by integrating over the 
Grassman variables, with the paramagnetic result \cite{fur:95}
\begin{equation}
\label{gg}
G(i\nu_n )\uI =\Bigl\langle \Bigl\{ G_0(i\nu_n)^{-1}\uI +\tJH \underline{\bs } \cdot \hm \Bigr\}^{-1} 
\Bigr\rangle_{\hm } 
=\ds\frac{G_0(i\nu_n)^{-1}}{G_0(i\nu_n)^{-2}-\tJH^2 }\uI,
\end{equation} 
where $\uI $ is the unity matrix in 2 x 2 spin space.  The average over the orientations 
$\hm $ of the local moment is generally given by
$\langle C(\hm )\rangle_{\hm }=\int d\Omega_{\hm } P(\hm )C(\hm )$,
where $P(\hm )\propto \Tr \Bigl( \exp (-\seff (\hm )\Bigr) $ is the
probability for the local moment to point in the $\hm $ direction.
Above $\TC $, $P(\hm )=1/4\pi $ is constant.  Consequently, the paramagnetic self-energy is given
by $\Sigma (i\nu_n ) = G_0(i\nu_n)^{-1}-G(i\nu_n)^{-1}=\tJH^2 G_0(i\nu_n)$. 
Expanded in powers of $\tJH $ and $G(i\nu_n)$, we find
\begin{equation}
\label{exp}
\Sigma (i\nu_n) = -\frac{1}{2G(i\nu_n)}+\sqrt{\ds\frac{1}{4G(i\nu_n)^2}+\tJH^2}
=\tJH^2 G(i\nu_n)-\tJH^4 G(i\nu_n)^3 +2\tJH^6 G(i\nu_n)^5 +\ldots . 
\end{equation}
On a Bethe lattice, these relations are closed by the analytic expression \cite{fur:95,geo:96} 
\begin{equation}
\label{sc}
\underline{G}_0(i\nu_n)^{-1}=z_n \uI  -\frac{W^2}{16}\underline{G}(i\nu_n),
\end{equation}
where $z_n=i\nu_n +\mu$ and $W$ is the full bandwidth of the non-interacting,
semicircular density-of-states.  We denote the full spin dependence for later use.

\begin{figure}
\includegraphics *[scale=0.5]{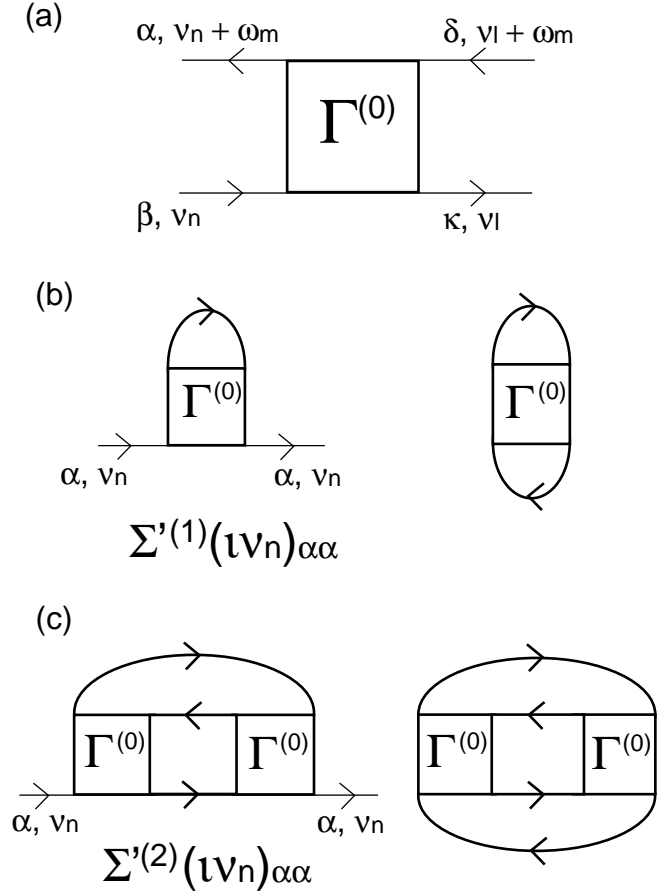}
\caption{
(a) The bare vertex function;  (b) and (c) Compact diagrams that contribute to
$\Phi $ for the electronic effective action $\seff'$ on the right 
with their associated self-energies on the left. 
}
\end{figure}

Diagrammatic perturbation theory is customarily formulated in terms of the bare vertex function 
$\G0 (l,n;m)^{\beta \alpha ;\delta \kappa }$ sketched in Fig.1(a) with $\omega_m=2m\pi T$.  
The bare vertex function may be associated with the 
two-particle interaction in the purely electronic effective action \cite{abr:63}
\begin{eqnarray}
\label{two}
&\seff' =-T\ds\sum_n\bc_{0\al }(i\nu_n) G_0(i\nu_n)^{-1} c_{0\alpha }(i\nu_n)\nonumber \\ & 
-\ds\frac{T^3}{4}\ds\sum_{l,n,m}\bc_{0\al }(i\nu_n+i\omega_m)c_{0\beta }(i\nu_n)
\G0 (l,n;m)^{\beta \al ; \delta \kappa }\bc_{0\kappa }(i\nu_l)c_{0\delta }(i\nu_l+i\omega_m).
\end{eqnarray}
Hence, the bare vertex function must satisfy the crossing symmetries $\G0 (l,l+m;n-l)^{\delta \alpha ;\beta \kappa }
=\G0 (n+m,n;l-n)^{\beta \kappa ;\delta \alpha } =-\G0 (l,n;m)^{\beta \al ;\delta \kappa }$.
There are two ways to calculate $\G0 (l,n;m)^{\beta \al ;\delta \kappa }$.
First, we can take the $\JH \rightarrow 0$ limit 
of the full irreducible vertex $\Gamma (l,n;m)^{\beta \alpha ; \delta \kappa }$ 
obtained from the Bethe-Salpeter equation for the magnetic susceptibility \cite{fis:03,fis:un}.
Alternatively, we can associate the lowest-order, $\JH^2$ contribution to the partition function
$Z=\langle \Tr \Bigl( \exp (-\seff (\hm )) \Bigr) \rangle_{\hm }$ 
with the contribution to the partition function $Z' =\Tr \Bigl( \exp (-\seff')\Bigr)$,
sketched as the compact diagram in Fig.1(b) (with internal lines given by the bare Green's 
functions $G_0(i\nu_n)_{\al \beta }$).  Both methods yield the same result:
\begin{equation}
\label{G0}
\G0 (l,n; m)^{\beta \al ;\delta \kappa }=\frac{1}{3}\beta \tJH^2 \Bigl\{
\bs_{\beta \al }\cdot \bs_{\delta \kappa }\delta_{m,0}-\bs_{\delta \al }\cdot \bs_{\beta \kappa }
\delta_{ln}\Bigr\},
\end{equation}
which satisfies the crossing symmetries.

However, replacing $\seff (\hm )$ by $\seff'$ produces an inequivalent theory \cite{ineq}.
For example, expanding $Z$ and $Z'$ in powers of $\JH $ yields the results
\begin{equation}
Z=Z_0\Bigl\{ 1 -\tJH^2 \sum_n G_0(i\nu_n)^2 +\frac{1}{2}\tJH^4 \sum_{l \ne n}G_0(i\nu_l)^2G_0(i\nu_n)^2+{\cal O}(\tJH^6) \Bigr\} ,
\end{equation}
\begin{equation}
Z'=Z_0\Bigl\{ 1 -\tJH^2 \sum_n G_0(i\nu_n)^2 +\frac{5}{6}\tJH^4 \sum_{l \ne n}G_0(i\nu_l)^2G_0(i\nu_n)^2+{\cal O}(\tJH^6) \Bigr\},
\end{equation}
which disagree to order $\tJH^4$.
Hence, it is not possible by averaging over the local moments to reduce the Hund's coupling to
an effective two-particle interaction.  In other words, the Hund's coupling produces 
fourth and higher-order electronic interactions that require higher-order
vertex functions in the electronic action.

A theory is usually said to be $\Phi $-derivable if a functional $\Phi (\{ \underline{G}(i\nu_n)\}) $, 
constructed from the sum of compact diagrams in terms of the full Green's functions and the 
bare vertex functions, can be found to satisfy the condition  
$\Sigma (i\nu_n)_{\al \beta }=\pt \Phi /\pt G(i\nu_n)_{\al \beta }$.
As discussed by Baym \cite{bay:62}, a $\Phi $-derivable theory may readily be shown to be 
thermodynamically consistent, meaning that thermodynamic properties can be evaluated 
either from the Green's function or from the partition function $Z$.
For a $\Phi $-derivable theory, the partition function $Z$ or free energy $-T\log Z$
may be constructed in terms of $\Phi $ from the relation
\begin{equation}
\label{z1}
-\log Z = \Phi -\sum_n \Tr \Bigl\{ \underline{\Sigma} (i\nu_n)\underline{G}(i\nu_n)\Bigr\}
+ \sum_n \Tr \log \Bigl\{ \underline{G}(i\nu_n)\Bigr\},
\end{equation}
which is stationary under variations of $\uG (i\nu_n)$.
Whereas Baym's original work was intended for systems of interacting Fermions and Bosons, 
the notion of $\Phi $-derivability has been extended to systems of interacting electrons and 
spins \cite{bic:87} and to disordered alloys \cite{lea:68}.  

From the discussion above, it is clear that even if it exists, $\Phi $ cannot be
constructed in terms of the bare vertex functions.  When the action contains only 
two-particle interactions such as for the Hubbard model, then the
first two terms in $\Phi $ are represented by the compact diagrams on the  
right-hand side of Figs.1(b) and (c) with the corresponding self-energies 
$\Sigma (i\nu_n)_{\al \al }=\pt \Phi /\pt G(i\nu_n)_{\al \al }$
sketched on the left-hand side.  Not surprisingly,
substituting our earlier expression for the bare vertex function produces
the correct first-order self-energy $\Sigma^{'(1)}(i\nu_n)=\tJH^2 G(i\nu_n)$ but the 
wrong second-order self-energy 
$\Sigma^{'(2)}(i\nu_n)=-(\tJH^4 /3)\Bigl\{ 2G(i\nu_n)\sum_lG(i\nu_l)^2+G(i\nu_n)^3\Bigr\}$.  Notice from 
Eq.(\ref{exp}) that the correct second-order self-energy $\Sigma^{(2)}(i\nu_n)=-\tJH^4 G(i\nu_n)^3$ 
does not involve a Matsubara summation.  Hence, the DMFT of the DE model is not $\Phi $-derivable 
in the strict diagrammatic sense stated above.

Despite the failure of a diagrammatic expansion in powers of $\underline{\Gamma }^{(0)}$, 
a functional $\Phi (\{ \uG (i\nu_n)\})$ can still be constructed 
to satisfy the condition
$\Sigma (i\nu_n)_{\al \al }= \pt \Phi /\pt G(i\nu_n)_{\al \al }$. 
Starting from Eq.(\ref{gg}) and Dyson's equation for the self-energy, we find that
$\pt \Sigma (i\nu_l)_{\al \al }/\pt G(i\nu_n)_{\beta \beta }=(\uK^{-1})_{ln}^{\al \beta }
+\delta_{ln}\delta_{\al \beta }G(i\nu_n)^{-2}$, where $\uK $ is the Jacobian 
\begin{equation}
K^{\al \beta }_{ln}= \ds\frac{\pt G(i\nu_n)_{\beta \beta }}{\pt [G_0(i\nu_l)_{\al \al }]^{-1}}
=-\delta_{ln}\ds\frac{1}{a_n^2}\Biggl\{ \ds\frac{2\tJH^2}{3}+b_n \delta_{\al \beta }\Biggr\} 
+\ds\frac{\tJH^2}{3a_la_n}\Bigl(2\delta_{\al \beta }-1\Bigr),
\end{equation}
with $a_n=G_0(i\nu_n)^{-2}-\tJH^2$ and $b_n=G_0(i\nu_n)^{-2}-\tJH^2/3$.
This Jacobian can be inverted with the general result
\begin{equation}
\label{jac}
\ds\frac{\pt \Sigma (i\nu_l)_{\al \al }}{\pt G(i\nu_n)_{\beta \beta }}
=-\delta_{ln}\ds\frac{\tJH^2 a_n^2}{3b_n}\Biggl\{ \ds\frac{2}{2a_n-3b_n} 
+\delta_{\al \beta }G_0(i\nu_n)^2\Biggr\}
-\ds\frac{\tJH^2 }{3-2\tJH^2\sum_r 1/b_r }\ds\frac{a_la_n}{b_lb_n}
\Bigl( 2\delta_{\al \beta }-1\Bigr).
\end{equation}
It can be shown \cite{fis:un} that the right-hand side equals 
$-T\,\Gamma (l,n;m=0)^{\al \al ;\beta \beta }$ where $\Gamma (l,n;m)^{\beta \alpha ; \delta \kappa }$ 
is the full irreducible vertex of the Bethe-Salpeter equation.  
The functional $\Phi $ must exist because the curl of the self-energy vanishes: 
$\pt \Sigma (i\nu_n)_{\al \al }/\pt G(i\nu_l)_{\beta \beta  } -
\pt \Sigma (i\nu_l)_{\beta \beta }/\pt G(i\nu_n)_{\al \al } =0$.

By construction, $\Phi^{(1)}$ (second order in $\JH $) is represented 
by the compact diagram in Fig.1(b) and is given in terms of the bare vertex function by 
\begin{eqnarray}
\label{p1}
\Phi^{(1)}& = -\ds\frac{T}{2}\sum_{l,r} \G0 (l,r;0)^{\al \al ;\beta \beta }
G(i\nu_l)_{\al \al }G(i\nu_r)_{\beta \beta }
=-\ds\frac{\tJH^2}{6}\biggl\{ \ds\sum_{l,n}G(i\nu_l)_{\al \al }
\nonumber \\ &
\times \Bigl( G(i\nu_n)_{\al \al }-G(i\nu_n)_{\bal \bal }\Bigr)
-\ds\sum_n G(i\nu_n)_{\al \al }^2-2\ds\sum_n G(i\nu_n)_{\al \al }G(i\nu_n)_{\bal \bal }\biggr\},
\end{eqnarray}
where $\bal $ is the opposite spin to $\al $.  After expanding 
and integrating Eq.(\ref{jac}) \cite{rcurv}, we find that $\Phi^{(2)}$ (fourth order in $\JH$) 
is given by 
\begin{eqnarray}
\label{p2}
\Phi^{(2)}& =\ds\frac{\tJH^4 }{9} \biggl\{ -\ds\frac{1}{4}
\ds\sum_n G(i\nu_n)_{\al \al }^4
-2\ds\sum_n G(i\nu_n)_{\al \al }^2 G(i\nu_n )_{\bal \bal }^2
-\ds\sum_{l,n,r} G(i\nu_r)_{\al \al }G(i\nu_r)_{\bal \bal } G(i\nu_l)_{\al \al }
\nonumber \\ &
\times \Bigl( G(i\nu_n )_{\al \al } -G(i\nu_n)_{\bal \bal }\Bigr)
+\ds\frac{2}{3}\ds\sum_{l,n} G(i\nu_n)_{\al \al }^3 \Bigl(
G(i\nu_l )_{\al \al }-G(i\nu_l)_{\bal \bal }\Bigr)\biggr\}.
\end{eqnarray}
Unlike $\Phi^{(1)}$, $\Phi^{(2)}$ cannot be represented by a compact diagram involving only the 
bare vertex functions.  So far, all of our results are valid for any lattice topology.  

We have verified the thermodynamic consistency of the DMFT by calculating the magnetic susceptibility
from both the Green's function and the partition function.  With a magnetic field $\bH =H\bz $ 
coupled to both the local moments and the electrons, the effective action becomes
\begin{equation}
\label{act2}
\seff (\hm ) =-T\sum_n\bc_{0\alpha  }(i\nu_n) \Bigl\{ G_0(i\nu_n)^{-1}_{\alpha \beta }
+ \Bigl(\tJH \hm +\frac{1}{2}H\bz \Bigr)\cdot \bs_{\alpha \beta }\Bigr\} c_{0\beta }(i\nu_n)-\beta HS m_z. 
\end{equation}
Parameterizing the bare inverse Green's function as $\underline{G}_0(i\nu_n )^{-1}=(z_n+R_n)\uI +Q_n\underline{\sigma }_z$
and using Eq.(\ref{sc}) for the full Green's function, we solve for $R_n$ and $Q_n$ on a Bethe lattice 
from the expression 
\begin{equation}
R_n\uI +Q_n\underline{\sigma }_z =-\frac{W^2}{16} \Bigl\langle \Bigl\{ (z_n+R_n)\uI +\Bigl(
\tJH \hm +(Q_n+H/2)\bz \Bigr)\cdot \underline{\bs }\Bigr\}^{-1} \Bigr\rangle_{\hm }. 
\end{equation} 
To linear order in the field, $R_n$ and $Q_n$ satisfy the implicit relations
\begin{equation}
R_n=-\frac{W^2}{16}\frac{z_n+R_n}{(z_n+R_n)^2-\tJH^2},
\end{equation}
\begin{equation}
Q_n=\frac{H(z_n+R_n) -2\tJH M_{lm }R_n}{2(z_n+2R_n)U_n}-\frac{H}{2},
\end{equation}
where
\begin{equation}
U_n = 1 -\frac{32\tJH^2}{3W^2}\frac{R_n^2}{(z_n+R_n)(z_n+2R_n)}.
\end{equation}
After integrating $\exp (-\seff (\hm ))$ over the Grassman variables, we find that the probability 
for the local moment to point along $\hm $ is
\begin{equation}
P(\hm ) \propto \exp \Biggl\{ \sum_n \log \Biggl( 1 - \frac{\tJH (2Q_n+H) m_z}{(z_n+R_n)^2-\tJH^2 }\Biggr)
+\beta HSm_z \Biggr\} \propto \exp (\beta J_{eff} M_{lm} m_z ),
\end{equation}
where the local-moment order parameter $M_{lm}=\langle m_z \rangle $ is solved from the condition 
$M_{lm}=J_{eff}M_{lm}\beta /3$.    
The electronic order parameter $M_{el}=2 \langle s_{0z}\rangle $ is obtained from the summation 
$M_{el} = -(32T/W^2)\sum_n Q_n$.  The total susceptibility is then given by the 
zero-field limit of $\chi = (SM_{lm }+M_{el}/2)/H$. 
To calculate the susceptibility from the partition function, we first expand $Z$ to second
order in $H$ and $M_{lm}$ and then use $\chi =(T/H)\partial \log Z/\partial H\vert_{H=0}$. 
The latter technique is formally equivalent to evaluating the susceptibility from the 
Bethe-Salpeter equation \cite{fis:03}.  

These two sets of calculations do indeed produce the same magnetic susceptibility, which may 
be written as
\begin{equation}
\label{sus}
\chi = \frac{1}{3T}\frac{\speff (T)^2}{1-(\tJH /W)^2 G_1(T)} +\frac{3T}{4W^2}\Bigl( G_1(T)
-G_2(T)\Bigr) +\frac{8\tJH^2 T}{W^4}G_1(T),
\end{equation}
\begin{equation}
\speff (T) = S +\frac{3\tJH T}{2W^2}\Bigl( G_1(T) -G_2(T)\Bigr),
\end{equation}
where the functions $G_1(T)$ and $G_2(T)$ are formally given by the Matsubara sums
\begin{equation}
G_1(T)=-\frac{32}{3}\sum_n \frac{R_n^2}{(z_n+R_n)(z_n+2R_n) U_n},
\end{equation} 
\begin{equation}
G_2(T)=-\frac{32}{3}\sum_n \frac{R_n}{(z_n+R_n)U_n}. 
\end{equation} 
The Curie temperature $\TC $ is solved from the condition $G_1(T_C)=(W/\tJH )^2$.

Previous results \cite{fis:03} in the $\JH \rightarrow \infty $ limit are reproduced \cite{limit} 
by taking $\mu = \sgn (p-1 )\tJH +\delta \mu$ where $\vert \delta \mu \vert \le W/(2\sqrt{2})$ and 
$p$ is the electron filling ($p=1$ means one electron per site).  The general expression
for the magnetic susceptibility shall be studied in a future publication.  We pause here to note 
that the effective spin $\speff (T)$ may be either larger or smaller than $S$ depending on the 
sign of the Hund's coupling $\JH $.  The temperature dependence of $\speff (T)$ and the 
deviation of $1-(\tJH /W)^2G_1(T)$ from $T-\TC$ are both caused by electronic correlations that
are absent in a local-moment system \cite{fis:03}.  The second and third sets of terms in Eq.(\ref{sus}) 
correspond to the Pauli susceptibility of the electrons.

Although $\Phi (\{ \uG (i\nu_n)\})$ has no simple diagrammatic interpretation, the 
existence of this functional means that we may still use Eq.(\ref{z1})
to establish the thermodynamic consistency of the DMFT of the
DE model.  Diagrammatics may be recovered for a more sophisticated model
where the classical local moments are replaced by fully quantum-mechanical 
operators and we introduce an additional propagator corresponding to those
local spins.  It may also be possible to develop a more complex diagrammatics 
for classical local spins in terms of higher-order vertex functions.

Finally, we note that whereas any conserving theory (in the sense of Baym and 
Kadanoff \cite{bay:62}) is thermodynamically consistent, it is not true that all 
thermodynamically consistent theories are conserving.  
Indeed, that is the case here since the DMFT violates the Ward identities associated 
with charge and spin conservation.

It is a pleasure to acknowledge helpful conversations with Dr. Gonzalo Alvarez and
Prof. Jim Freericks.  This research was sponsored by the U.S. Department of Energy under contract 
DE-AC05-00OR22725 with Oak Ridge National Laboratory, managed by UT-Battelle, LLC and
by the National Science Foundation under Grant No. EPS-0132289 (ND EPSCoR) and
DMR-0312680.


\begin{references}
\bibitem{ineq} The electronic effective action $\seff'$ would still produce an inequivalent theory
even if the three-dimensional local spins $S\hm $ were replaced by Ising spins $\pm S\bz$.
\bibitem{rcurv} Due to the term containing $\Bigl\{ 1-(2/3)\tJH^2 \sum_l \Bigl(
G_0(i\nu_l)^{-2}-\tJH^2/3\Bigr)^{-1}\Bigr\}^{-1}$
in the irreducible vertex function $\underline{\Gamma }$, the radius of convergence of the 
expansion in powers of $\tJH $ is of order $T$.  In the limit $\JH \rightarrow \infty $, 
$\underline{\Gamma }$ depends on the cutoff $n_{\chi }$ of the Matsubara sum \cite{fis:03}.
\bibitem{limit} In the $\JH \rightarrow \infty $ limit, $G_1(T)\rightarrow (W/\tJH)^2 F_1(T)$
and $G_2(T)\rightarrow (W^2/3\tJH T)\bigl\{ \vert p-1 \vert -1 -2\, \sgn (p-1)F_2(T) \bigr\} $, where
$F_1(T)$ and $F_2(T)$ were defined in Ref.\cite{fis:03}.

\suppressfloats
\end{references}
\end{document}